\documentstyle[aps,epsfig,amsmath,amssymb]{revtex}

\draft

\begin{document}

\title{Finite size effects on the collective oscillations of a trapped
Bose gas}
   
\author{F.~Zambelli and S.~Stringari}
\address{Dipartimento  di Fisica, Universit\`a di Trento,}
\address{and Istituto Nazionale per la Fisica della Materia, I-38050 Povo,
Italy}
\date{\today}

\maketitle

\begin{abstract}
Sum rules are used to calculate the corrections to the hydrodynamic 
collective frequencies of a harmonically  trapped Bose gas at $T=0$,
due to finite size effects. We show that, with logarithmic accuracy,
the relative corrections behave like $(a_{\rm ho}/R)^4\log(R/a_{\rm
ho})$, where $R$ is the Thomas-Fermi radius of the condensate and
$a_{\rm ho}$ is the oscillator length. Results are given for both
spherical and axi-symmetric trapping.  
\end{abstract}

  
\begin{section}{Introduction}
\label{intro}

After the experimental realization of Bose-Einstein condensation (BEC)
in atomic trapped alkali gases \cite{BEC} the study of their dynamical
properties has become  an important subject of experimental and theoretical
research (see \cite{rmp} and references therein). The measurements 
of the collective frequencies have confirmed the validity of the mean
field approach \cite{rmp,SS,GPosc} in describing the dynamical behaviour
of a trapped Bose-Einstein condensed gas at zero temperature, properly 
accounting for the important role played by two-body interaction and
by Bose statistics. The high precision of these frequency measurements is
stimulating new perspectives in the study of finer  effects like  
the temperature dependence of the collective frequencies, the
occurrence of beyond mean field corrections \cite{LPSS}, {\em etc}. 
In this perspective it is important to have a full analytical control 
of the values of the collective frequencies predicted by the mean
field scheme at $T=0$, including finite size corrections
which are neglected in the Thomas-Fermi limit. 
The purpose of the present work  is to calculate these 
corrections  in the framework of the mean field Gross-Pitaevskii
theory, employing a sum rule approach \cite{Lippa}. In general 
sum rules  provide only an upper bound to the frequency of the lowest modes.
In this paper we will show that, due to the harmonic nature of the
confinement, the first corrections to the Thomas-Fermi limit can be 
calculated with logarithmic accuracy using this method. We will show that the
corrections are fixed by the kinetic energy term which, in these
trapped Bose-Einstein condensed systems, plays the role of an effective 
surface tension \cite{Usama}. 
\end{section}

\begin{section}{Sum rules and collective modes}
\label{sec1}

The usefulness of the sum rule technique in studying the dynamical
behaviour of trapped Bose-Einstein condensates has been already
discussed in several papers (see, for example, \cite{rmp,Francesca}). 
This method, typical of many-body physics, is based on the calculation of
the energy-weighted moments  $m_p=\int\!d\omega\,
S_{\cal F}(\omega)\,(\hslash\omega)^p$  of the dynamic structure function
associated with a given operator ${\cal F}$. The dynamic structure
function, at zero temperature, takes the form
\begin{equation}
S_{\cal F}(\omega)=\sum_n|\langle n|{\cal F}
|0\rangle|^2\delta(\hslash\omega-\hslash\omega_{n0})\;,
\label{strength}
\end{equation}
where $\hbar\omega_{n0}=\hbar\omega_n-\hbar\omega_0$ 
is the excitation energy of the eigenstate $|n\rangle$ of the
Hamiltonian, whose ground state is $|0\rangle$.
At $T=0$ the quantity $S_{\cal F}(\omega)$ is related to the dynamic response
function of the system relative to ${\cal F}$ through the expression 
\cite{Pines} 
\begin{equation}
\chi_{\cal F}(\omega)=
\int_{-\infty}^{\infty}\!\!\!\!d\omega\left[\frac{S_{\cal F}(\omega^{\prime})}
{\omega-\omega^{\prime}+i\eta}-\frac{S_{\cal F}(\omega^{\prime})}
{\omega+\omega^{\prime}+i\eta}\right]\;. 
\label{chi}
\end{equation}
In this work we make the following choice for the Hamiltonian 
\begin{equation}
H=\frac{1}{2M}\sum_{i=1}^Np_i^2+\frac{M}{2}
\sum_{i=1}^N\big(\omega_x^2x_i^2+\omega_y^2y_i^2+\omega_z^2z_i^2\big)
+g\sum_{i<j}\delta({\mathbf r}_i-{\mathbf r}_j)\;, 
\label{H}
\end{equation} 
where the parameter $g=4\pi\hslash^2a/M$ is the
strength of the two-body contact interaction, fixed by the $s$-wave
scattering length $a$. 
Ratios between different sum rules provide rigorous upper bounds 
\begin{equation}
\hbar \omega_{\rm lowest} \le \sqrt{m_{p+2}/m_p}
\label{ineqlowest}
\end{equation}
to the energy of the lowest state excited by the operator ${\cal F}$. 
These bounds coincide with $\omega_{\rm lowest}$ only if the  strength
of the operator ${\cal F}$, defined by Eq. (\ref{strength}), is
exhausted by a single mode.

By using the closure relationship involving the eigenstates $|n\rangle$, the
energy-weighted and the cubic moments can be easily rewritten  in the
form of sum rules as follows
\begin{align}
m_1&=\frac{1}{2}\langle\big[{\cal F}^{\dagger},[H,{\cal F}]\big]\rangle
\label{m1}\\
m_3&=\frac{1}{2}\langle \big[[{\cal F}^{\dagger},H],[H,[H,{\cal
F}]]\big]\rangle\;,
\label{m3}
\end{align}
where the averages are taken over the ground state of the system
which, in the mean field scheme, is provided by the stationary
solution of the Gross-Pitaevskii equation \cite{LP,GP}.

Let us first consider the simplest situation where the harmonic confinement
is isotropic, {\em i.e.} $\omega_x=\omega_y=\omega_z=\omega_{\rm ho}$. 
In this case the calculation of the above moments is straightforward
(see Appendix \ref{appendixA}) if the operator ${\cal F}$ coincides
with the monopole 
\begin{equation}
{\cal M}=\sum_{i=1}^Nr_i^2
\label{M}
\end{equation}
where $r_i^2=x_i^2+y_i^2+z_i^2$, or quadrupole 
\begin{equation}
{\cal Q}=\sum_{i=1}^Nx_iy_i
\label{Q}
\end{equation}
operators, yielding the following results for the ratio $\sqrt{m_3/m_1}$
\begin{align}
\omega_{{\cal M}_{3,1}}&=\sqrt{5}\,\omega_{\rm ho}\left[1-\frac{E_{\rm kin}}
{5E_{\rm ho}}\right]^{1/2}
\label{fM31}\\
\omega_{{\cal Q}_{3,1}}&=\sqrt{2}\,
\omega_{\rm ho}\left[1+\frac{E_{\rm kin\perp}}
{E_{\rm ho\perp}}\right]^{1/2}\;.
\label{fQ31}
\end{align}
In Eqs. (\ref{fM31})-(\ref{fQ31}) 
$E_{\rm kin}=\langle\sum_{i=1}^Np_i^2\rangle/2M$ and 
$E_{\rm ho}=M\omega_{\rm ho}^2\langle\sum_{i=1}^Nr_i^2\rangle/2$ are the
expectation values of the kinetic and harmonic oscillator energies
respectively,
$E_{\rm kin\perp}=\langle\sum_{i=1}^N(p_{x_i}^2+p_{y_i}^2)\rangle/2M=
2\,E_{\rm kin}/3$ and 
$E_{\rm ho\perp}=M\omega_{\rm ho}^2\langle
\sum_{i=1}^N(x_i^2+y_i^2)\rangle/2=2\,E_{\rm ho}/3$ are their
corresponding radial contributions.
In deriving Eq. (\ref{fM31}) we have also used the virial identity \cite{rmp} 
\begin{equation}
2E_{\rm kin}-2E_{\rm ho}+3E_{\rm int}=0
\label{virial}
\end{equation}
relating in an exact way the various contributions to the ground state 
energy
\begin{equation}
E=E_{\rm kin}+E_{\rm ho} +E_{\rm int}
\label{E}
\end{equation}
where $E_{\rm int}=g\langle\sum_{i<j}\delta ({\mathbf r}_i-{\mathbf
r}_j)\rangle$ is the average value of the mean field interaction energy.

Another possibility for calculating the monopole and the quadrupole
frequencies makes use of  the ratio $\sqrt{m_1/m_{-1}}$, which
involves the inverse energy weighted moment. In general the following 
inequality holds
\begin{equation}
\sqrt{m_1/m_{-1}} \le \sqrt{m_3/m_1}
\label{inequality}
\end{equation}
showing, by comparison with (\ref{ineqlowest}), 
that the lowest excitation frequency is in general better approximated 
by the ratio  $\sqrt{m_1/m_{-1}}$ than by $\sqrt{m_3/m_{1}}$.
Differently from $m_1$ and $m_3$, the moment $m_{-1}$  cannot be
written in terms of commutators, but is related to the static response
function via the following general relationship (compressibility sum
rule \cite{Pines})
\begin{equation}
\chi_{\cal F}(0)=-2\sum_n\frac{|\langle n|{\cal F}|0\rangle
|^2}{\hslash\omega_{n0}}=-2m_{-1}\;,
\label{chistatica}
\end{equation}
which provides a natural rule for calculating $m_{-1}$ through the explicit 
determination of the polarization $\langle{\cal F}+{\cal
F}^{\dagger}\rangle=\gamma\chi_{\cal F}$ induced in the system by an external
static field interacting through the Hamiltonian 
$\gamma({\cal F} + {\cal F}^{\dagger})$.

We present in Appendix \ref{appendixB} the details of the derivation
and report here only the final results \cite{rmp,tesi}:   
\begin{align}
\omega_{{\cal M}_{1,-1}}&=2\,
\omega_{\rm ho}\left[\langle r^2\rangle\bigg/\left(
\langle r^2\rangle -\frac{N}{2}\frac{\partial}{\partial N}
\langle r^2\rangle\right)\right]^{1/2}
\label{fM1-1}\\
\omega_{{\cal Q}_{1,-1}}&=
\sqrt{2}\,\omega_{\rm ho}\lim_{\varepsilon\to
0}\left(\frac{\partial\delta}{\partial\varepsilon}\right)^{-1/2}\;,
\label{fQ1-1}
\end{align}
where 
$\varepsilon=(\omega_x^2-\omega_y^2)/(\omega_x^2+\omega_y^2)$ and
$\delta=\langle y^2-x^2\rangle /\langle y^2+x^2\rangle$ are,
respectively, the deformation of the confining potential and of the
condensate in the $x$-$y$ plane.
Equations (\ref{fM1-1}) and (\ref{fQ1-1}) show that the monopole
frequency $\omega_{{\cal M}_{1,-1}}$ is determined by the
$N$-dependence of the square radius 
$\langle r^2\rangle=\langle\sum_{i=1}^Nr_i^2\rangle /N$, 
while the quadrupole by the dependence of the deformation $\delta$
of the cloud on the  deformation $\varepsilon$ of the trap. 
In (\ref{fQ1-1}) we have explicitly taken the limit of a symmetric trap. 

It is worth pointing out that the results (\ref{fM31})-(\ref{fQ31}) and
(\ref{fM1-1})-(\ref{fQ1-1}) are valid in general in the framework of the 
mean field Gross-Pitaevskii theory and apply, in particular,
to both repulsive and attractive forces \cite{nota1}. 
In particular they reproduce the correct values for the frequencies in
both the relevant limits of the non interacting gas, 
where $E_{\rm kin}=E_{\rm ho}$, and of the Thomas-Fermi approximation
where the kinetic energy is negligible with respect to $E_{\rm ho}$
\cite{Baym}.   
In the former case both the monopole and quadrupole frequencies
approach the value $2\,\omega_{\rm ho}$, while in the latter 
they tend to the analytic hydrodynamic values  $\sqrt{5}\,\omega_{\rm ho}$ 
and $\sqrt{2}\,\omega_{\rm ho}$ for the monopole and  quadrupole
oscillations respectively, as  predicted in Ref. \cite{SS}. 
This feature allows us to conclude  that in the hydrodynamic regime a single
mode exhausts the sum rules $m_3$, $m_1$ and $m_{-1}$. 
In the following section we will show that, with logarithmic accuracy,
the same is true also if one includes the leading finite size
corrections to the Thomas-Fermi limit.
\end{section}

\begin{section}{Finite-size corrections in spherically symmetric systems}
\label{sec2}

In the Thomas-Fermi regime the asymptotic behaviour of the kinetic energy
per particle of a harmonically isotropic Bose gas at $T=0$ is given by
\cite{Franco,Lundh}
\begin{equation}
\frac{E_{\rm kin}}{N}=\frac{5}{2}
\frac{\hslash^2}{MR^2}\log{\bigg(\frac{R}{Ca_{\rm ho}}\bigg)}
\label{EkinTF}
\end{equation}
where 
\begin{equation}
R=\left(\!15\,\frac{Na}{a_{\rm ho}}\right)^{1/5}a_{\rm ho}
\label{RTF}
\end{equation}
is the Thomas-Fermi radius of the condensate, with $a_{\rm
ho}=\sqrt{\hbar/M\omega_{\rm ho}}$, and $C\simeq 1.3$ is a numerical factor. 
Such a behaviour arises from a careful analysis of the shape of the
order parameter near the border of the condensate, where the
Thomas-Fermi approximation fails and the wave function of the condensate
should be calculated with proper accuracy. 
By using the virial identity (\ref{virial}), and the usual definition
$\mu=\partial E/\partial N$ for the chemical potential
one can easily show that the first corrections to $E_{\rm ho}/N$ and $E_{\rm
int}/N$ beyond the Thomas-Fermi approximation are given by: 
\begin{align}
\frac{E_{\rm ho}}{N}&=\frac{3}{7}\mu_{\rm TF} +
\frac{\hslash^2}{MR^2}\left[
\log{\bigg(\frac{R}{Ca_{\rm ho}}\bigg)}+\frac{3}{8}\right] 
\label{EhoTF}\\
\frac{E_{\rm int}}{N}&=\frac{2}{7}\mu_{\rm TF} -
\frac{\hslash^2}{MR^2}\left[
\log{\bigg(\frac{R}{Ca_{\rm ho}}\bigg)}-\frac{1}{4}\right]\;,
\label{EintTF}
\end{align}
where 
\begin{equation}
\mu_{\rm TF}=\frac{1}{2}M\omega_{\rm ho}^2R^2
\label{muTF}
\end{equation}
is the Thomas-Fermi chemical potential. Note that in deriving the
above equations we have consistently taken into account all the terms of
order $1/R^2$, which renormalize the constant $C$ inside the $\log$
argument.
Eqs. (\ref{EhoTF})-(\ref{EintTF}) were also derived in Ref. \cite{FF}
by employing a boundary layer theory. 
 
From Eqs. (\ref{EkinTF}) and (\ref{EhoTF}) one finds that the first
corrections to the ratio $E_{\rm kin}/E_{\rm ho}$ can be written as
\begin{equation}
\frac{E_{\rm kin}}{E_{\rm ho}}=
\frac{35}{3}\left(\frac{a_{\rm ho}}{R}\right)^4\times
\log{\left(\frac{R}{Ca_{\rm ho}}\right)}
\label{ratio}
\end{equation}
and becomes smaller and smaller in the large $R$, Thomas-Fermi
limit. Equations (\ref{EhoTF}) and (\ref{EkinTF}) also reveal that
the kinetic energy plays the role of an effective surface energy,
providing the first correction to the
total energy (\ref{E}) in the Thomas-Fermi regime.

By using the above results it is immediate to determine the
finite-size corrections to the hydrodynamic value of the frequencies
predicted by Eqs. (\ref{fM31}) and (\ref{fQ31}). It is interesting
also to evaluate these corrections starting from Eqs. (\ref{fM1-1}) and
(\ref{fQ1-1}). Appendix \ref{appendixC} is devoted to the technical
details of this calculation, which shows that ratios $\omega_{3,1}$
and $\omega_{1,-1}$ do coincide for both the monopole and the
quadrupole modes, apart from a different value of the coefficient inside
the $\log{}$ argument.  
The results for the collective frequencies, including the first
corrections due to finite size effects, take the form
\begin{align}
\lim_{R\gg a_{\rm ho}}\omega_{\cal M}&
=\sqrt{5}\,\omega_{\rm ho}\left[1-\frac{7}{6}
\left(\frac{a_{\rm ho}}{R}\right)^4
\log{\left(\frac{R}{C_{\cal M}\,a_{\rm ho}}\right)}
\right]
\label{fMcorr}\\
\lim_{R\gg a_{\rm ho}}\omega_{\cal Q}&
=\sqrt{2}\,\omega_{\rm ho}\left[1+\frac{35}{6}
\left(\frac{a_{\rm ho}}{R}\right)^4
\log{\left(\frac{R}{C_{\cal Q}\,a_{\rm ho}}\right)}
\right]\;,
\label{fQcorr}
\end{align}
with $C_{\cal M}=C\,e^{-1/8}$ and $C_{\cal Q}=C\,e^{3/5}$ for
$\omega_{1,-1}$, while $C_{\cal M}=C_{\cal Q}=C$ if one uses the estimate
$\omega_{3,1}$.
Notice that our results differ from the ones found in Ref. \cite{FF}, 
where corrections behaving like $(a_{\rm ho}/R)^4$ (without 
the logarithmic factor) were obtained by employing a different 
procedure based on a perturbative solution for the excited state wave function.

The physical role played by the kinetic energy and the $R$ dependence 
of the finite size corrections can be understood in terms of the
classical picture developed in Ref. \cite{Usama} for the surface modes of a
Bose condensed gas confined by a linear potential. In this case the resulting 
dispersion relation takes the form $\omega^2 = Fq/m$ plus  $q^4\log q$ 
corrections originating from the kinetic energy term, where $q$ is
the wave vector of the surface wave. In this equation 
$F=m\omega_{ho}^2R$ is the harmonic force evaluated at the
Thomas-Fermi radius $R$. 
The correspondence with our result (\ref{fQcorr}) for the surface quadrupole
mode is achieved by imposing the discreteness condition
$q \sim  1/R$ to the wave vector.

It is also worth noticing the different sign of the finite size
corrections entering the monopole and quadrupole frequencies.
Furthermore the correction for the quadrupole is about $5$ times larger than
for the monopole.

It is finally interesting to compare the finite size corrections  predicted 
by Eqs. (\ref{fMcorr}) and (\ref{fQcorr}) with the ones due to beyond
mean field effects  \cite{LPSS}. 
These are fixed by the gas parameter
\begin{equation}
a^3n(0)=\frac{15^{2/5}}{8\pi}\left(N^{1/6}\frac{a}{a_{\rm ho}}\right)^{12/5}
\label{bmf}
\end{equation}
calculated at the central density $n(0)$ of the gas. These effects are not
included in the mean field Gross-Pitaevskii theory and take into
account the changes in the equation of state originating from quantum
correlation effects according to the theory of Lee-Yang-Huang
\cite{LYH}. The dependence of the gas parameter (\ref{bmf}) on the relevant
physical variables ($N$, $a$ and $a_{\rm ho}$) is quite different
with respect to the one exhibited by the ratio (\ref{ratio}).
Furthermore in Ref. \cite{LPSS} it was found that only the monopole oscillation
is affected by beyond mean field effects. 
This should introduce enough flexibility to distinguish, in the study of the
collective frequencies,  between finite size and  beyond mean field effects.
First experimental efforts in this direction were made on $^{85}$Rb
where the value of the scattering length can be tuned, thereby
amplifying the effects of quantum correlations \cite{noteeric}. 
\end{section} 
 
\begin{section}{Results for deformed traps} 
\label{sec3}

Until now our analysis was restricted to the case of spherical
traps. It is interesting, especially from the experimental
viewpoint, to extend the above results to the case of deformed external
potentials. We will focus on harmonic traps having
axial symmetry ($\omega_x=\omega_y=\omega_{\perp}$ and
$\omega_z=\lambda\,\omega_{\perp}$, the parameter $\lambda$
characterizing the shape of the trap).    
Having proven in the previous section that the accuracy of the sum
rule approach is logarithmic, we will present in the following only
the corrections to the estimate $\sqrt{m_3/m_1}$, whose calculation is
straightforward. 

We are interested in particular in the study of the collective excitations
described by the ($\ell =2$, $m=2$) quadrupole operator (\ref{Q}) and
by the operator
\begin{equation}
{\cal B}=\sum_{i=1}^N\big(\alpha r_i^2+\beta z_i^2\big)\;,
\label{m=0mode}
\end{equation}
which accounts for the coupling between the $\ell =2$, $m=0$ and the
the $\ell =0$ excitations. 
The parameters $\alpha$ and $\beta$ fix the relative weight of the two modes. 
Moreover one can show that by treating them as variational parameters
within a sum rule calculation, one recovers the correct dispersion for
the two decoupled modes in Thomas-Fermi regime, first obtained in
Ref. \cite{SS} by solving the linearized hydrodynamic equations of superfluids.
   
Let us first discuss the $m=2$ mode. By using the results
(\ref{m1quadrupole}) and (\ref{m3quadrupole}) reported in Appendix 
\ref{appendixA} it is straightforward to find the result
\begin{equation}
\lim_{R_{\perp}\gg a_{\perp}}\omega_{\cal Q}=
\sqrt{2}\,\omega_{\perp}\left[1+\frac{7}{4}
\frac{E_{{\rm kin}\,\perp}}{N\mu_{\rm TF}}\right]\;,
\label{omegaQdef}
\end{equation} 
where the Thomas-Fermi chemical potential is the natural
generalization of Eq. (\ref{muTF})
\begin{equation} 
\mu_{\rm TF}=\frac{1}{2}M\omega_{\perp}^2R_{\perp}^2\;,
\label{muTF2}
\end{equation}
with the radial size of the condensate given by 
\begin{equation}
R_{\perp}=\left(\!15\,\lambda\frac{Na}{a_{\perp}}\right)^{2/5}a_{\perp}
\label{Rperp}
\end{equation}
and $a_{\perp}=\sqrt{\hbar/M\omega_{\perp}}$.

For the $m=0$ mode, by using Eqs. (\ref{m1B}) and (\ref{m3B}), we find
\begin{equation}
\omega^2(m=0)=
\frac{1}{NM}\frac{\alpha^2(10\,E_{\rm ho}-2\,E_{\rm
kin})+2\alpha\beta(10\,E_{{\rm ho}\,z}-2\,E_{{\rm
kin}\,z})+\beta^2(6\,E_{{\rm ho}\,z}+2\,E_{{\rm kin}\,z})}
{\alpha^2\langle r^2\rangle
+2\alpha\beta\langle z^2\rangle +\beta^2\langle
z^2\rangle}\;,
\label{m=031}
\end{equation}
for the ratio $(\hbar\omega)^2=m_3/m_1$,
where $E_{{\rm kin}\,z}=\langle\sum_{i=1}^N p_{z_{i}}^2
\rangle/2M$ and $E_{{\rm ho}\,z}=M\omega_z^2\langle\sum_{i=1}^N{z_{i}}^2
\rangle/2$ are the axial contributions to $E_{\rm kin}$ and $E_{\rm
ho}$ respectively.
By imposing that the ratio (\ref{m=031}) be stationary with respect to
the variation of $\alpha$ (or $\beta$) one can determine the decoupled
eigenfrequencies. We present here only the explicit results in the
simpler cases of strongly deformed traps. In the limit $\lambda\ll 1$
(cigar) one finds that the lowest of the two frequencies of
Eq. (\ref{m=031}) reads
\begin{equation}
\lim_{R_{\perp}\gg a_{\perp}}
\omega\,(m=0)=\sqrt{\frac{5}{2}}\,\omega_z\left[1+\frac{7}{20}
\frac{E_{{\rm kin}\,\perp}}{N\mu_{\rm TF}}\right]\;.
\label{m=0lowcigar} 
\end{equation}
Conversely when $\lambda\gg 1$ (disk) one gets 
\begin{equation}
\lim_{R_{\perp}\gg a_{\perp}}
\omega\,(m=0)=\sqrt{\frac{10}{3}}\,\omega_{\perp}\left[1+
\frac{14}{15}\frac{E_{{\rm kin}\,z}}{N\mu_{\rm TF}}\right]\;.
\label{m=0lowcake}
\end{equation}

According to Eqs. (\ref{omegaQdef}), (\ref{m=0lowcigar}) and
(\ref{m=0lowcake}), in order to make quantitative estimates of the
finite size corrections one needs an explicit expression for 
$E_{{\rm kin}\,\perp}$ and $E_{{\rm kin}\,z}$. We have calculated these
quantities by employing the method developed in Ref. \cite{Lundh}. 
In Appendix \ref{appendixD} we report the results, which explicitly
depend on the deformation of the trap.
\end{section}

\begin{section}{Conclusions}
\label{end}

In this paper we have used a sum rule approach to calculate the
frequency of the monopole and quadrupole modes of dilute Bose gas
confined by a harmonic trap at $T=0$. We have shown that with
logarithmic accuracy the sum rules are exhausted by a single mode not
only in Thomas-Fermi regime but also if one includes the first
corrections due to the finite size of the could.
We have provided an explicit expression for the frequency shift
of the monopole and quadrupole modes due to finite size effects both
for spherical and deformed traps.
\end{section}

\acknowledgments
Useful discussions with L.P.~Pitaevskii are acknowledged.

\appendix
\section{Results for the energy weighted and the cubic sum rules}
\label{appendixA}

The $m_1$ and $m_3$ sum rules can be calculated by carrying out the
commutators of Eqs. (\ref{m1})-(\ref{m3}) with the mean field
Hamiltonian (\ref{H}). The results for $m_{-1}$ are instead derived in
Appendix \ref{appendixB}.

For the monopole operator (\ref{M}) one finds:
\begin{align}
m_{1}^{{\cal M}}&=\frac{2\hbar^2}{M}N\langle r^2\rangle
\label{m1monopole}\\
m_{3}^{{\cal M}}&=\frac{2\hbar^4}{M^2}\big(4E_{\rm kin}+4E_{\rm ho}+9E_{\rm
int}\big)\;.
\label{m3monopole}
\end{align}

For the quadrupole operator (\ref{Q}) the results read:
\begin{align}
m_{1}^{{\cal Q}}&=\frac{\hbar^2}{2M}N\langle x^2+y^2\rangle
\label{m1quadrupole}\\
m_{3}^{{\cal Q}}&=\frac{2\hbar^4}{M^2}\big(E_{{\rm kin}\,\perp}+E_{{\rm
ho}\,{\perp}}\big)\;.
\label{m3quadrupole}
\end{align}

For the operator (\ref{m=0mode}) we find:
\begin{align}
m_1^{{\cal B}}&=\frac{2\hbar^2}{M}N\big(\alpha^2\langle r^2\rangle
+2\alpha\beta\langle z^2\rangle +\beta^2\langle z^2\rangle\big)
\label{m1B}\\
m_3^{{\cal B}}&=\frac{2\hbar^2}{M^2}\big(\alpha^2(10E_{\rm ho}-2E_{\rm
kin})+2\alpha\beta(10E_{{\rm ho}\,z}-2E_{{\rm
kin}\,z})+\beta^2(6E_{{\rm ho}\,z}+2E_{{\rm kin}\,z})\big)\;.
\label{m3B}
\end{align}

\section{The inverse energy weighted sum rule}
\label{appendixB}
The aim of this appendix is to determine the inverse energy weighted
sum rule relative to the monopole and quadrupole operators (\ref{M})
and (\ref{Q}) when the Hamiltonian of the system is given by
Eq. (\ref{H}), with $\omega_x=\omega_y=\omega_z=\omega_{\rm ho}$.
The procedure is based on the determination of the static
response function through Eq. (\ref{chistatica}).

{\em Monopole}:
in this case the problem is equivalent to finding the static response
when the system is perturbed by an external field
proportional to the monopole operator (\ref{M}), which can be
described by the Hamiltonian:
\begin{equation}
H_{\rm pert}=\gamma\sum_{i=1}^Nr_i^2\;,
\label{HpertM}
\end{equation}
where $\gamma$ is the strength of the perturbation.
Notice that adding such a perturbation to the Hamiltonian (\ref{H})
has simply the effect of renormalizing the frequency of the external
confinement, which now becomes
\begin{equation}
\omega_{\gamma}=\sqrt{\omega_{\rm ho}^2+\frac{2\gamma}{M}}\;.
\label{omegaeff}
\end{equation} 
As a consequence, the average square radius will now depend on
$\gamma$.
 
By definition, the static polarizability is given by 
\begin{equation}
\chi_{\cal M}(\omega=0)=N\left(\frac{\partial\langle r^2\rangle_{\gamma}}
{\partial\gamma}\right)_{\gamma=0}\;.
\label{chistatica2}
\end{equation}  
In order to evaluate (\ref{chistatica2}), one should exploit the $\gamma$
dependence of $\langle r^2\rangle_{\gamma}$ in an explicit way.
To this purpose it is convenient to write the stationary
Gross-Pitaevskii equation in a suitable form, by rescaling the
lengths, the energies and the order parameter as suggested in Ref.
\cite{GPnum}:
\begin{align}
{\mathbf r}&=a_{\gamma}\,{\mathbf r}_1\\
E&=\hbar\omega_{\gamma}\,E_1\\
\psi({\mathbf r})&=\sqrt{\frac{N}{a_{\gamma}^3}}\,\psi_1({\mathbf r}_1)\;,
\end{align}
with $\psi_1({\mathbf r}_1)$ normalized to $1$.
According to such notation the time-independent GP equation for
$\psi_1({\mathbf r}_1)$ reads
\begin{equation}
[-\nabla_1^2+r_1^2+u_1|\psi({\mathbf r}_1)|^2]
\psi_1({\mathbf r}_1)=2\mu_1\psi_1({\mathbf r}_1)\;,
\label{GPred}
\end{equation}
where $u_1=8\pi\,Na/a_{\gamma}$, $a_{\gamma}=\sqrt{\hbar/M\omega_{\gamma}}$ 
and $\mu=\hbar\omega_{\gamma}\mu_1$.
Consequently one can write the square radius of the system as 
\begin{equation}
\langle r^2\rangle_{\gamma}=a_{\gamma}^2\langle r^2_1\rangle_{\gamma}
=a_{\gamma}^2\int\!d{\mathbf r}_1 \,r_1^2 \,|\psi_1({\mathbf r}_1)|^2\;.
\label{r2split}
\end{equation}
Note that $\langle r^2_1\rangle_{\gamma}$ exhibits a dependence on
$\gamma$, since the Gross-Pitaevskii equation, in the form
(\ref{GPred}), depends on $\gamma$ through  the mean field term
proportional to $u_1$. The derivative of $\langle r^2\rangle_{\gamma}$
with respect to $\gamma$ can then be expressed in terms of the
derivative with respect to $u_1$ or $N$.

By using Eqs. (\ref{r2split}) and (\ref{chistatica2}) one finally finds
\begin{align}
\chi_{\cal M}(\omega=0)&=N\left[\langle r^2_1\rangle_{\gamma}
\left(\frac{\partial a_{\gamma}^2}{\partial\gamma}\right)
+a_{\gamma}^2\left(\frac{\partial\langle r^2_1\rangle_{\gamma}}
{\partial\gamma}\right)
\right]_{\gamma=0}\nonumber\\
&=N\left[-\frac{a_{\gamma}^2}{M\omega_{\rm ho}^2}\left(
\langle r^2_1\rangle_{\gamma}-\frac{N}{2}
\frac{\partial\langle r^2_1\rangle_{\gamma}}{\partial N}
\right)\right]_{\gamma =0}\;,
\label{chistaticaM}
\end{align}
which, combined with Eq. (\ref{chistatica}), gives  the relevant result
\begin{equation}
m_{-1}^{\cal M}=\frac{N}{2M\omega_{\rm ho}^2}\left(\langle
r^2\rangle-\frac{N}{2}\frac{\partial\langle r^2\rangle}{\partial N}\right)\;.
\label{m-1M}
\end{equation}

{\em Quadrupole}:
in this case the idea is that in the presence of a harmonic potential deformed
in the $x$-$y$ plane ($\omega_x\neq\omega_y$) the quadrupole operator
(\ref{Q}) can be written also in the form
\begin{equation}
{\cal Q}=\frac{1}{iM\hbar}\frac{[H,L_z]}{(\omega_x^2-\omega_y^2)}\;,
\label{Q2}
\end{equation}
where 
\begin{equation}
L_z=\sum_{i=1}^N(x_ip_{y_i}-y_ip_{x_i})
\label{Lz}
\end{equation}
is the third component of the angular momentum operator.
By using the relationship (\ref{chistatica}) combined with the
expression (\ref{Q2}) for the quadrupole operator, one easily shows
that $m_{-1}^{\cal Q}$ can be written as
\begin{equation}
m_{-1}^{\cal Q}=\frac{1}{2}
\frac{\langle [L_z,{\cal Q}]\rangle}{i\hbar M(\omega_x^2-\omega_y^2)}\;.
\label{m-1Q0}
\end{equation}
From Eq. (\ref{m-1Q0}), taking the spherical limit, one immediately
finds the result
\begin{equation}
m_{-1}^{\cal Q}=\frac{N}{4M\omega_{\rm ho}^2}
\lim_{\varepsilon\to 0}\,\langle
x^2+y^2\rangle\,\left(\frac{\partial\delta}{\partial\varepsilon}\right)\;, 
\label{m-1Q}
\end{equation}
where the deformation of the could $\delta$ and the one of the trap
$\varepsilon$ have been introduced in Sec. \ref{sec2}.

\section{Comparison between $\hbox{\mbox{\boldmath $\omega_{3,1}$}}$ 
 and $\hbox{\mbox{\boldmath $\omega_{1,-1}$}}$ }
\label{appendixC}
We report here the full derivation of the finite size corrections
relative to both monopole and the quadrupole modes in the isotropic case.

{\em Monopole}: It is immediate to derive the corrections to $\omega_{3,1}$
by using the Eq. (\ref{fM31}) combined with the Eq. (\ref{ratio}).
One finds the result
\begin{equation}
\omega_{{\cal M}_{3,1}}
=\sqrt{5}\,\omega_{\rm ho}\left[1-\frac{7}{6}
\left(\frac{a_{\rm ho}}{R}\right)^4
\log{\left(\frac{R}{Ca_{\rm ho}}\right)}
\right]\;.
\label{fMcorr31}
\end{equation}

It is quite simple also to calculate the finite size corrections to 
$\omega_{1,-1}$ starting from Eq. (\ref{fM1-1}). One has to use the identity
\begin{equation}
\langle r^2\rangle=\frac{2}{M\omega_{\rm ho}^2}\frac{E_{\rm ho}}{N}\;,
\label{r2}
\end{equation} 
with the harmonic oscillator energy given by Eq. (\ref{EhoTF}). 
Note that the $N$-dependence of the mean square radius $\langle
r^2\rangle$ originates from both the Thomas-Fermi chemical potential
(\ref{muTF}) and from the correction behaving like 
$1/R^2\log(R/a_{\rm ho})$, where $R$ is given in
Eq. (\ref{RTF}).
By carrying out carefully the calculation and taking into account all
the terms of order $1/R^4$ , one finally finds
\begin{equation}
\omega_{{\cal M}_{1,-1}}
=\sqrt{5}\,\omega_{\rm ho}\left[1-\frac{7}{6}
\left(\frac{a_{\rm ho}}{R}\right)^4
\log{\left(\frac{R\,e^{1/8}}{Ca_{\rm ho}}\right)}
\right]\;,
\label{fMcorr1-1}
\end{equation}
which coincides with the result (\ref{fMcorr31}) for $\omega_{3,1}$
within logarithmic accuracy.
Note that  the inequality (\ref{inequality}) is correctly fulfilled by
Eqs. (\ref{fMcorr31}) and (\ref{fMcorr1-1}), the relative difference
between the two estimates becoming smaller than $10^{-4}$ for $R >
6\,a_{\rm ho}$. 

{\em Quadrupole}: the calculation of the corrections starting from the ratio
$\sqrt{m_3/m_1}$ is immediate if one combines Eqs. (\ref{fQ31}) and
(\ref{ratio}). This yields the result
\begin{equation}
\omega_{{\cal Q}_{3,1}}
=\sqrt{2}\,\omega_{\rm ho}\left[1+\frac{35}{6}
\left(\frac{a_{\rm ho}}{R}\right)^4
\log{\left(\frac{R}{Ca_{\rm ho}}\right)}
\right]\;.
\label{fQcorr31}
\end{equation} 

The calculation of the corrections to $\omega_{1,-1}$ is less trivial
since it requires the explicit knowledge of 
the dependence of the condensate deformation $\delta$ on $\varepsilon$.
To this purpose we determine the finite size corrections to
the square radii $\langle x^2\rangle=2\,E_{{\rm ho}\,x}/NM\omega_x^2$ and 
$\langle y^2\rangle=2\,E_{{\rm ho}\,y}/NM\omega_y^2$ by using the
virial relationship (\ref{virial}) written for the $x$ direction
\begin{equation}
E_{{\rm kin}\,x}-E_{{\rm ho}\,x}+\frac{1}{2}E_{\rm int}=0\;,
\label{virial1D}
\end{equation} 
the analogous equations holding for $y$ and $z$.
One then finds
\begin{align}
\langle y^2-x^2\rangle&=\frac{1}{NM}
\frac{\omega_x^2+\omega_y^2}{\omega_x^2\omega_y^2}
\left[E_{\rm int}(\varepsilon)\,
\varepsilon+
(1+\varepsilon)E_{{\rm kin}\,y}(\varepsilon)-
(1-\varepsilon)E_{{\rm kin}\,x}(\varepsilon)\right]
\label{num_delta}\\
\langle y^2+x^2\rangle&=\frac{1}{NM}
\frac{\omega_x^2+\omega_y^2}{\omega_x^2\omega_y^2}
\left[E_{\rm int}(\varepsilon)
+(1+\varepsilon)E_{{\rm kin}\,y}(\varepsilon)+
(1-\varepsilon)E_{{\rm kin}\,x}(\varepsilon)\right]\;,
\label{den_delta}
\end{align} 
where we have explicitly indicated the dependence of $E_{\rm int}$, 
$E_{{\rm kin}\,x}$ and $E_{{\rm kin}\,y}$ on the trap deformation.
By expanding the above equations with respect to $\varepsilon$ one can
easily determine the limit entering Eq. (\ref{fQ1-1})
getting, to the lowest order,
\begin{equation}
\omega_{{\cal Q}_{1,-1}}=\sqrt{2}\,\omega_{\rm ho}
\lim_{\varepsilon\to 0}
\left(\frac{\partial\delta}{\partial\varepsilon}\right)^{-1/2}=
\sqrt{2}\,\omega_{\rm ho}\left[
1+\frac{7}{4}\frac{E_{{\rm kin}\,x}^{\prime}(0)-E_{{\rm
kin}\,y}^{\prime}(0)}{N\mu_{\rm TF}}\right]\;,
\label{fQcorr1-1_1}
\end{equation}
with $E_{{\rm kin}\,x}^{\prime}(0)=(\partial E_{{\rm
kin}\,x}/\partial\varepsilon)_{\varepsilon =0}$ and analogously for
$y$.
In order to estimate the difference $E_{{\rm kin}\,x}^{\prime}(0)-E_{{\rm
kin}\,y}^{\prime}(0)$ entering the last equation, we can use the
results reported in the next Appendix \ref{appendixD} for the
different contributions to the kinetic energy in Thomas-Fermi regime.
By setting $\lambda=\omega_y/\omega_x$ and using the relationship
$\varepsilon\simeq (1-\lambda) $ holding for small $\varepsilon$, one
finds that
\begin{equation}
E_{{\rm kin}\,x}^{\prime}(0)-E_{{\rm kin}\,y}^{\prime}(0)=
\frac{2}{3}E_{\rm kin}-N\frac{\hslash^2}{MR^2}\;,
\label{corrections}
\end{equation} 
which combined with Eq. (\ref{fQcorr1-1_1}) finally gives 
\begin{equation}
\omega_{{\cal Q}_{1,-1}}=\sqrt{2}\,\omega_{\rm ho}\left[1+\frac{35}{6}
\left(\frac{a_{\rm ho}}{R}\right)^4
\log{\left(\frac{R}{Ca_{\rm ho}\,e^{3/5}}\right)}\right]\;.
\label{fQcorr1-1}
\end{equation}
It is worth pointing out that, with logarithmic accuracy,
Eq. (\ref{fQcorr1-1}) coincides with the result (\ref{fQcorr31})
derived for $\omega_{3,1}$. 

\section{Kinetic energy in Thomas-Fermi regime}
\label{appendixD}

By explicitly calculating the various contributions $E_{{\rm
kin}\,\perp}$ and $E_{{\rm kin}\,z}$ of the total kinetic energy using
the method developed in Ref. \cite{Lundh} we find the following results:

\noindent
{\em Disk shaped trap}, $\lambda > 1$
\begin{align}
\frac{E_{\rm kin,\perp}}{N}&=\frac{\hbar^2}{2MR_{\perp}^2}\left[
\frac{5}{2}\log{\left(\frac{2}{\lambda^{2/3}}\frac{R_{\perp}}{\delta_0}
\right)}-
\frac{5}{2}\left(1+\frac{1}{3}\frac{1}{\lambda^2-1}\right)
\frac{\arctan{\big(\sqrt{\lambda^2-1}}\big)}{\sqrt{\lambda^2-1}}+\frac{5}{18}
\frac{1+2\lambda^2}{\lambda^2-1}  
\right]
\label{ekinperp1}\\
\frac{E_{\rm kin,z}}{N}&=\frac{\hbar^2}{2MR_{\perp}^2}\left[
\frac{5}{4}\lambda^2
\log{\left(\frac{2}{\lambda^{2/3}}\frac{R_{\perp}}{\delta_0}\right)}+
\frac{5}{6}\left(1+\frac{1}{\lambda^2-1}\right)
\frac{\arctan{\big(\sqrt{\lambda^2-1}}\big)}{\sqrt{\lambda^2-1}}-\frac{5}{18}
\frac{1+2\lambda^2}{\lambda^2-1}-\frac{5}{18}-\frac{5}{9}\lambda^2
\right]
\label{ekinz1}
\end{align}

\noindent
{\em Cigar shaped trap}, $\lambda < 1$
\begin{align}
\frac{E_{\rm kin,\perp}}{N}&=\frac{\hbar^2}{2MR_{\perp}^2}\left[
\frac{5}{2}\log{\left(\frac{2}{\lambda^{2/3}}\frac{R_{\perp}}{\delta_0}
\right)}-
\frac{5}{4}\left(1-\frac{1}{3}\frac{1}{1-\lambda^2}
\right)\frac{1}{\sqrt{1-\lambda^2}}
\log{\left(\frac{1+\sqrt{1-\lambda^2}}{1-\sqrt{1-\lambda^2}}\right)}
-\frac{5}{18}\frac{1+2\lambda^2}{1-\lambda^2}
\right]
\label{ekinperp2}\\
\frac{E_{\rm kin,z}}{N}&=\frac{\hbar^2}{2MR_{\perp}^2}\left[
\frac{5}{4}\lambda^2
\log{\left(\frac{2}{\lambda^{2/3}}\frac{R_{\perp}}{\delta_0}\right)}+
\frac{5}{12}\left(1-\frac{1}{1-\lambda^2}
\right)\frac{1}{\sqrt{1-\lambda^2}}
\log{\left(\frac{1+\sqrt{1-\lambda^2}}{1-\sqrt{1-\lambda^2}}\right)}
+\frac{5}{18}\frac{1+2\lambda^2}{1-\lambda^2}
-\frac{5}{18}-\frac{5}{9}\lambda^2
\right]\;.
\label{ekinz2}
\end{align}
In Eqs. (\ref{ekinperp1})-(\ref{ekinz2})
$\delta_0$ represents the surface thickness at $z=0$.
Its dependence on the relevant parameters of the system is \cite{Lundh}
\begin{equation}
\delta_0=\frac{1.776}{2^{1/3}}\left(\!
15\,\lambda\frac{Na}{a_{\perp}}\right)^{-1/15}\!\!a_{\perp}\;.
\end{equation}
The above results are consistent with the results of Ref. \cite{Lundh}
for the total kinetic energy $E_{\rm kin}=E_{{\rm kin}\,\perp}+E_{{\rm
kin}\,z}$, which reduces to (\ref{EkinTF}) in the isotropic case $\lambda=1$.


\begin{references}
\bibitem{nota1}
In the  case of attractive forces it can be shown that the estimate 
(\ref{fM1-1}) for the monopole frequency explicitly reveals the trend of the
system to collapse when the number of particles approaches a critical
threshold. Indeed the monopole mode becomes softer and softer and
exactly vanishes at this threshold.
\end{references}
\end{document}